\def\eqref#1{equation~\ref{#1}}
\def\1{\bm{1}}

\def\rmX{{\mathbf{X}}}

\def\vb{{\bm{b}}}

\def\vf{{\bm{f}}}

\def\vh{{\bm{h}}}

\def\vt{{\bm{t}}}

\def\vv{{\bm{v}}}

\def\vx{{\bm{x}}}

\DeclareMathAlphabet{\mathsfit}{\encodingdefault}{\sfdefault}{m}{sl}
\SetMathAlphabet{\mathsfit}{bold}{\encodingdefault}{\sfdefault}{bx}{n}

\def\gB{{\mathcal{B}}}

\def\gE{{\mathcal{E}}}

\def\gG{{\mathcal{G}}}

\def\gM{{\mathcal{M}}}
\def\gN{{\mathcal{N}}}

\def\sR{{\mathbb{R}}}

\documentclass{article}
\usepackage[utf8]{inputenc} % allow utf-8 input
\usepackage[T1]{fontenc}    % use 8-bit T1 fonts
\usepackage{hyperref}       % hyperlinks
\usepackage{url}            % simple URL typesetting
\usepackage{booktabs}       % professional-quality tables
\usepackage{amsfonts}       % blackboard math symbols
\usepackage{nicefrac}       % compact symbols for 1/2, etc.
\usepackage{microtype}      % microtypography
\usepackage{xcolor}         % colors
\usepackage{pifont}   
\usepackage[numbers]{natbib}
\usepackage{cancel}
\usepackage{microtype}
\usepackage{graphicx}
\usepackage{subfigure}
\usepackage{booktabs} % for professional tables
\usepackage{arydshln}
\usepackage{hyperref}
\usepackage{url}
\usepackage{graphicx}
\usepackage{makecell}
\usepackage{multirow}
\usepackage{enumitem}
\usepackage{bbm}
\usepackage{color,xcolor}
\usepackage{booktabs}
\usepackage{lineno}
\usepackage{array}
\usepackage{ulem}
\usepackage{makecell}
\usepackage{multicol}
\usepackage{amsmath,amsthm,amsfonts,amssymb,bm}
\usepackage{stfloats}
\usepackage[font=small,labelfont=bf]{caption}
\usepackage{tabulary}
\usepackage{algorithm}
\usepackage{algorithmicx}
\usepackage{algpseudocode}
\usepackage{bbding}
\usepackage{amsmath}
\usepackage{amsfonts}
\usepackage{graphicx} % 图片支持
\usepackage{wrapfig} % wrapfig宏包
\newlist{compactenum}{enumerate}{4}
\setlist[compactenum,1]{nolistsep}
\newcolumntype{C}[1]{>{\centering\arraybackslash}p{#1}}
\newcolumntype{L}[1]{>{\arraybackslash}p{#1}}

\DeclareMathOperator*{\argmax}{arg\,max}

\algnewcommand{\algorithmicendif}{}
\algnewcommand{\algorithmicforend}{}

\algtext*{EndIf}
\algtext*{EndFor}

\usepackage[preprint]{neurips_2023}

\usepackage[utf8]{inputenc} % allow utf-8 input
\usepackage[T1]{fontenc}    % use 8-bit T1 fonts
\usepackage{hyperref}       % hyperlinks
\usepackage{url}            % simple URL typesetting
\usepackage{booktabs}       % professional-quality tables
\usepackage{amsfonts}       % blackboard math symbols
\usepackage{nicefrac}       % compact symbols for 1/2, etc.
\usepackage{microtype}      % microtypography
\usepackage{xcolor}         % colors
\usepackage{wrapfig}
\usepackage{ulem}
\newcommand\com[1]{\text{\textcolor{gray}{#1}}}
\usepackage{colortbl}
\definecolor{Highlight}{rgb}{0.89,0.89,0.94}
\newcommand{\chl}{\cellcolor{Highlight}}
\definecolor{RelativeGain}{HTML}{E0F1F7}
\newcommand{\crg}{\cellcolor{RelativeGain}}

\definecolor{textbf}{HTML}{00AEF0}

\title{FoldToken2: Learning compact, invariant and generative protein structure language }

\author{Zhangyang Gao$^{\dagger}$, Cheng Tan $^{\dagger}$, Stan Z. Li$^{*}$\\
AI Lab, Research Center for Industries of the Future, Westlake University \\
\texttt{\{gaozhangyang, tancheng, Stan.ZQ.Li\}@westlake.edu.cn}\\
\thanks{$^{\dagger}$Equal Contribution, $^{*}$Corresponding Author.}
}
\makeatletter
\def\thanks#1{\protected@xdef\@thanks{\@thanks
        \protect\footnotetext{#1}}}
\makeatother

\begin{document}
\maketitle

\begin{abstract}
  The equivalent nature of 3D coordinates has posed long term challenges in protein structure representation learning, alignment, and generation. Can we create a compact and invariant language that equivalently represents protein structures? Towards this goal, we propose FoldToken2 to transfer equivariant structures into discrete tokens, while maintaining the recoverability of the original structures. From FoldToken1 to FoldToken2, we improve three key components: (1) invariant structure encoder, (2) vector-quantized compressor, and (3) equivalent structure decoder. We evaluate FoldToken2 on the protein structure reconstruction task and show that it outperforms previous FoldToken1 by 20\% in TMScore and 81\% in RMSD. FoldToken2 probably be the first method that works well on both single-chain and multi-chain protein structures quantization. We believe that FoldToken2 will inspire further improvement in protein structure representation learning, structure alignment, and structure generation tasks.
\end{abstract}

\section{Introduction}
\begin{quote}
    % "Struggling in the dark, find the light, then share to the community."
    "SE-(3) structure should not be special and difficult. Let's lower the barrier."
    \begin{flushright}
        -- Our Goal
    \end{flushright}
\end{quote}

Protein structure modelling plays a foundational role in computational biology and have attracted increasing attention in machine learning. Due to the SE-(3) equivariant nature, encoding \cite{zhang2022protein,fan2022continuous,gao2022pifold, gao2023kw, gao2024proteininvbench} and generating \cite{jumper2021highly, baek2021accurate, lin2023evolutionary, tan2024cross, abramson2024accurate} protein structure never to be trivial, which requires the special design targeted to protein structures. For example, PiFold \cite{gao2022pifold} proposes the invariant featurizer to encode structure patterns, and AlphaFold2 \cite{jumper2021highly} design frame-based model to generate equivariant 3D coordinates. While nuemerious innovations have been proposed in designing the protein structure models, the structures data itself remains the SE-(3) nature. \textit{Can we transform equivariant structures into a invariant form, and then use the existing NLP/CV models to encode and generate the structures?}

We introce FoldToken2, a novel method to transform SE-(3) structures into invariant representations. The key insight is to create a compact invariant latent space that preserve the structure information via self-reconstruction. After pretraining, the invariant latent representation can serve as prototype of the equivariant structures that is editable in the latent space. Akin to image and text, we also introduce a vector quantization module to discretize the latent space to create a SE-(3) invariant language. Taking the invariant latent embedding or language as input, we can use the existing CV or NLP models \cite{brown2020language, dosovitskiy2020image,gao2024graph,rombach2022high} to encode and generate the protein structures. FoldToken2 contains three key components: (1) invariant encoder, (2) vector quantization module, and (3) equivariant decoder.

A frame-based GNN (BlockGAT) is used for encoding equivariant structures as invariant embeddings. FoldToken1 \cite{gao2023vqpl,gao2024foldtoken} represent backbone structures as bond and torsion angles, lacking the ability to capture the 3D dependencies. As a remedy, FoldToken2 represent residues as block like AlphaFold2 \cite{jumper2021highly}, with a efficient and powerful graph neural network BlockGAT \cite{gao2024uniif}.  The BlockGAT contains a simplified featurizer for capturing the informative 3D dependencies and a optimized graph network module for learning high-level representations. The sparse graph attention mechanism make it much more efficient than the SE-(3) transformers, which is important for large-scale pre-training. 

A optimized vector quantization module (T-SoftCVQ) is used to quantize the continuous embeddings into discrete tokens, termed fold language. There are great success in applying vector quantization to image and video modelling, however, few attempt has tried to apply it to protein structure modelling  \cite{gao2023vqpl, gao2024foldtoken}. To make advanced sequence models, such as BERT \cite{devlin2018bert} and GPT \cite{brown2020language}, to be powerful structure learner, quantizing continuous embeddings into discrete tokens is crucial. Based on FoldToken1 \cite{gao2023vqpl,gao2024foldtoken}, we further improve the vector quantization module (SoftCVQ) by proposing new temperature annealing and encoding strategy, leading to T-SoftCVQ. Thanks to the elaborate design, the reconstruction results consistently outperform previous VQ methods.  

A conditional SE-(3) decoder is proposed for structure generation. The decoder takes discrete tokens as conditional features, and generates the 3D coordinates via equivariant graph message passing. Each SE-(3) layer include a BlockGAT and a plug-and-play SE-(3) module; By stacking multiple SE-(3) BlockGATs, we iteratively refine the protein structure from Gaussian noise conditioned on discrete tokens. During reconstruction, the BlockGAT extracts pairwise residue interactions online, and SE-(3) module uses frame-level message passing to refine the local frames.

% We propose a conditional SE-(3) BlockGAT to enhance structure generation. Notabely, we propose a plug-and-play SE-(3) module for structure generation. By adding the SE-(3) module to the BlockGAT, the model can update the residue structures via equivariant graph message passing, resulting in SE-(3) BlockGAT. We stack multiple SE-(3) BlockGATs to iteratively generate the protein structure from Gaussian noise conditioned on the invariant representations. During reconstruction, protein structures are repeatly refined by the SE-(3) module, and the featurizer update residue interactions online.

We evaluate the reconstruction performance of FoldToken2 on both single-chain and multi-chain settings. In single chain reconstruction, following FoldToken1 \cite{gao2023vqpl,gao2024foldtoken} which may be the first to tokenizing single-chain, FoldToken2 significantly improves the reconstruction performance on both TMScore and RMSD by 20\% and 81\%, respectively. In addition, FoldToken2 extends the idea to multi-chain protein structure reconstruction, and also lead to promising results. We believe that FoldToken2 will inspire further improvement in protein structure representation learning, structure alignment, and structure generation tasks.

\section{Method}
\subsection{Overall Framework}
\begin{figure*}[h]
    \centering
    \includegraphics[width=5.5in]{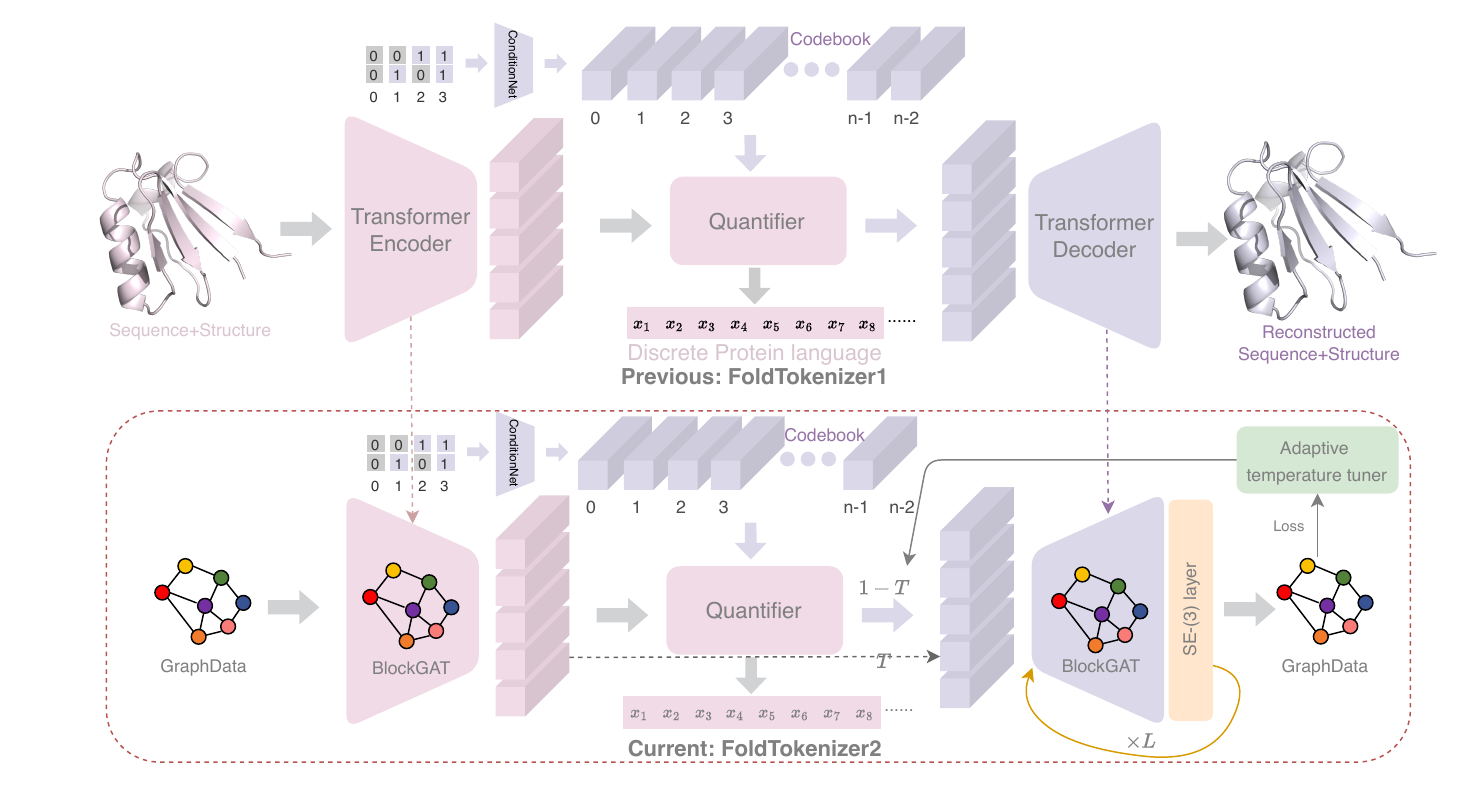}
    \caption{The overall framework of FoldTokenizer2, which contains contains encoder, quantifier, and decoder. In FoldToken2, we use BlockGAT to encoder protein structures as invariant embeddings, SoftCVQ to quantize the embeddings into discrete tokens, and SE-(3) layer to recover the protein structures iteratively.}
    \vspace{-5mm}
    \label{fig:framework}
 \end{figure*}

As shown in Fig.\ref{fig:framework}, the overall frame work keeps the same as FoldToken1 \cite{gao2023vqpl,gao2024foldtoken}, including encoder, quantifier and decoder. FoldToken2 comprehensively improves the each module to enhance the reconstruction performance, which are summarized as follows:

\begin{enumerate}
    \item \textbf{Data Form}: Changing from the angle-based representation to coordinate-based one.
    \item \textbf{Backbone}: We replace the transformer backbone with new GNN, named BlockGAT.
    \item  \textbf{VQ}: We introduce teacher-guided temperature annealing  strategy.
    \item \textbf{Generator}: We propose novel SE-(3) layer to refine the protein structures iteratively.
\end{enumerate}

\subsection{Invariant Graph Encoder}
Due to the rotation and translation equivariant nature, the same protein may have different coordinates records, posing a challenge of learning compact invariant representations for the same protein. Previous work \cite{gao2022pifold, jing2020learning, dauparas2022robust,gao2024uniif} has shown that the invariant featurizer can encode the invariant structure patterns, and we follow the same road: representing the protein structures as a graph consisting of invariant node and edge features. Then we use the BlockGAT \cite{gao2024uniif} to learn high-level representations.

\paragraph{Frame-based Block Graph.} 
Given a protein $\gM = \{\gB_s\}_{s=1}^n$ containing $n$ blocks, where each block represent a amino acid, we build the block graph $\gG(\{\gB_s\}_{s=1}^n, \gE)$ using kNN algorithm. In the block graph, the $s$-th node is represented as $\gB_s = (T_s, \vf_s)$, and the edge between $(s,t)$ is represented as $\gB_{st} = (T_{st}, \vf_{st})$. $T_s$ and $T_{st} = T_s^{-1} \circ T_t$ are the local frames of the $s$-th and the relative transform between the $s$-th and $t$-th blocks, respectively. $\vf_s$ and $\vf_{st}$ are the node and edge features.

\paragraph{BlockGAT Encoder.} We use the BlockGAT \cite{gao2024uniif} layer $f_{\theta}$  to learn block-level representations:
\begin{equation}
    \vf_s^{(l+1)}, \vf_{st}^{(l+1)} \leftarrow \texttt{BlockGATs}(\vf_s^{(l)},\vf_{st}^{(l)}| T_s, T_{st}, \gE)
\end{equation}
where $\vf_s^{(l)}$ and $\vf_{st}^{(l)}$ represent the input node and edge features of the $l$-th layer. $T_s=(R_s, \vt_s)$ is the local frame of the $s$-th block, and $T_{st} = T_s^{-1} \circ T_t = (R_{st}, \vt_{st})$ is the relative transform between the $s$-th and $t$-th blocks. $T_s, T_{st}, \vf_s^{(0)}$ and $\vf_{st}^{(0)}$ are initialized from the ground truth structures using the invariant featurizer proposed in \cite{gao2024uniif}.

\subsection{Quantifier}
Following FoldToken1 \cite{gao2023vqpl,gao2024foldtoken}, we use the SoftCVQ to quantize the invariant embeddings into discrete tokens, termed fold language. Instead of projecting continual embeddings into discrete tokens, SoftCVQ maps pre-defined binary number ($\vb_j$) into continuous token embeddings $\vv_j$ and then conduct soft alignment between the token embeddings $\vv_j$ and latent embeddings $\vh_s$. 

Given the decimal integer $z$ and the codebook size $m$, we represent $z$ as a binary vector $\vb_j$ with length $\log_2(m)$. For example, if $m=4$, we have $\vb_1=[0,0], \vb_2=[0,1], \vb_3=[1,0], \vb_4=[1,1]$. The quantization operation transforms continuous embeddings $\vh_s$ into discrete tokens $z$:
\begin{equation}
    \begin{cases}
       a_{sj} &= \frac{\exp{(\vf_s^T \vv_j/T)}}{\sum_{j=1}^m \exp{(\vf_s^T \vv_j/T)}}\\
       z &= \argmax_{j} a_{sj}
    \end{cases}
    \label{eq:quantifier}
\end{equation}
where $\vv_j$ is the $j$-th token embedding, generated by a conditional network, which also server as the de-quantization operation:
\begin{equation}
    \begin{cases}
       \vb_s &= \text{Bit}(z, \log_2(m))\\
       \hat{\vf}_s &= \text{ConditionNet}(\vb_s)
    \end{cases}
\end{equation}
In Eq.~\ref{eq:quantifier}, the "argmax" is non-differentiable, and we use a soft approximation during training:
\begin{equation}
    \begin{cases}
       \hat{\vf}_i &= \sum_{j=1}^m a_{ij} \vv_j\\
       \vb_j &= \text{Bit}(j, \log_2(m))\\
       \vv_j &= \text{ConditionNet}(\vb_j)
    \end{cases}
\end{equation}
The temperature parameter $T$ controls the softness of the attention query operation. When $T$ is large, the attention weights will be close to uniform; otherwise, the attention weights will be close to one-hot. During training, the temperature is annealed from 1.0 to 1e-8. The $\text{ConditionNet}: \sR^{\log_2(m)} \rightarrow \sR^{d}$ is a MLP. If the codebook size is $2^{16}$, the MLP projects 65536 16-dimension boolvectors into 65536 $d$-dimension vectors. Considering the gradient of the softmax operation:
\begin{equation}
    \begin{cases}
        [s_1, s_2, \cdots, s_k] = \text{Softmax}(z_1/T, z_2/T, \cdots, z_k/T)\\
        \frac{\partial s_i}{\partial z_j} = \frac{1}{T} s_i \cdot (\mathbbm{1}(i=j)-s_j)
    \end{cases}
\end{equation}
when $T$ tend to zero, the gradient explodes. To avoid this, we frozen the encoder and VQ module when the temperature is  lower than 1e-5. The temperature scheduler is:
\begin{equation}
    \begin{cases}
        x = 2 \pi (\frac{t\%T}{T}) - \pi\\
        T = \max(10^{-8}, \frac{1+\cos(x)}{2})  \times \beta
    \end{cases}
\end{equation}

To accelerate model convergence, we introduce:
\begin{itemize}
    \item \textbf{Teacher-Guided VQ}: We randomly copy the encoder embedding $\vf_s$ as the quantized feature $\hat{\vf}_s$ in a probability of $T$, termed teacher-guided vector quanternion.
    \item \textbf{Adaptive tuner}: We adaptively adjust the temperature scale $\beta$ based on training loss.
\end{itemize}
The adaptive temperature tuner is defined as:
\begin{equation}
    \beta =
    \begin{cases}
        1.0, \text{if } 0.1<\mathcal{L}\\
        0.05, \text{if } 0.05< \mathcal{L} \leq 0.1\\
        0.01, \text{if } 0.02< \mathcal{L} \leq 0.05\\
        0.0001, \text{if } \mathcal{L} \leq 0.02
    \end{cases}
\end{equation}

\subsection{Equivariant Graph Decoder}
Generating the protein structures conditioned on invariant representations poses significant challenges in computing efficiency. For example, training well-known AlphaFold2 from scratch takes 128 TPUv3 cores for 11 days \cite{wang2022helixfold}; OpenFold takes 50000 GPU hours for training \cite{ahdritz2024openfold}. In this work, we propose an efficiency plug-and-play SE3Layer that could be added to any GNN layer for structure prediction. Thanks to the simplified module of SE3Layer and BlockGAT with sparse attention, we can train the model over the whole PDB dataset in 1 day using 8 NVIDIA-A100s.

\paragraph{SE-(3) Frame Passing Layer.} We introduce frame-level message passing, which updates the local frame of the $s$-th block by aggregating the relative rotation $R_s$ and translation $\vt_s$ from its neighbors:
\begin{equation}
    \begin{cases}
        \text{vec}{(R_s)} = \sum_{j \in \gN_s} a_{sj}^r \text{vec}{(R_{sj})}\\
        R_{s} \leftarrow \text{Quat2Rot} \circ \text{Norm} \circ \text{MLP}^{9\rightarrow 4}(\text{vec}{(R_s)}) \quad \com{Normalize quanternion}\\
        \vt_s = \sum_{j \in \gN_s} a_{sj}^t \vt_{sj}
    \end{cases}
\end{equation}
where  $a_{sj}^r$ and $a_{sj}^t$ are the rotation and translation weights, and $\gN_s$ is the neighbors of the $s$-th block. $\text{vec}{(\cdot)}$ flattens $3\times3$ matrix to $9$-dimensional vector. $\text{MLP}^{9\rightarrow 4}(\cdot)$ maps the 9-dim rotation matrix to 4-dim quaternion, and $\text{Norm}(\cdot)$ normalize the quaternion to ensure it represents a valid rotation. $\text{Quat2Rot}(\cdot)$ is the quaternion to rotation function. We further introduce the details as follows:

% $q$ indicates the 4-dimensional quanternion vector.

\begin{equation}
    \begin{cases}
        w^r_{st}, w^t_{st} = \sigma(\text{MLP}(\vf_{st})) \\
        \text{vec}{(R_{st})} \leftarrow w^r_{st} \text{vec}{(R_{st})}+ (1-w^r_{st}) \text{MLP}^{d \rightarrow 9}(\vf_{st})\\ 
        \vt_{st} \leftarrow w^t_{st} \vt_{st} + (1-w^t_{st}) \text{MLP}^{d\rightarrow 3}(\vf_{st})\\
        a^r_{st}, a^t_{st} = \text{Softmax}(\text{MLP}^{d\rightarrow 1}(\vf_{st}))
    \end{cases}
    \label{se3}
\end{equation}
where $w^r_{st}$ and $w^t_{st}$ are the updating weights for rotation and translation, $a^r_{st}$ and $a^t_{st}$ are the attention weights. The propose SE-(3) layer could be add to any graph neural network for local frame updating.

\paragraph{Iterative Refinement} We propose a new module named SE-(3) BlockGAT by adding a SE-(3) layer to BlockGAT. We stack multi-layer SE-(3) BlockGAT to iteratively refine the structures:

\begin{equation}
    \begin{cases}
        \vf_s^{(l+1)}, \vf_{st}^{(l+1)} = \text{BlockGAT}^{(l)}(\vf_s^{(l)}, \vf_{st}^{(l)})\\
        T_{st}^{(l)} = T_s^{-1} \circ T_t \\
        T_s^{(l+1)} = \text{SE3Layer}(\texttt{sg}(T_{st}^{(l)}), \vf_{st}^{(l+1)})
    \end{cases}
\end{equation}
where $\texttt{sg}(\cdot)$ is the stop-gradient operation, and $\text{SE3Layer}(\cdot)$ is the SE-(3) layer described in Eq.\ref{se3}. Given the predicted local frame $T_s^{(l)}$, we can obtain the 3D coordinates by:
\begin{equation}
    \begin{cases}
        \vh_s = \text{MLP}(\vf_s^{(l)})\\
        \vx_s = T_s^{(l)} \circ \vh_s
    \end{cases}
\end{equation}

\subsection{Reconstruction Loss}
Inspired by Chroma \cite{ingraham2023illuminating}, we use multiple losses to train the model. The overall loss is:
\begin{equation}
    \mathcal{L} =  \mathcal{L}_{global} + \mathcal{L}_{fragment} + \mathcal{L}_{pair} + \mathcal{L}_{neighbor} + \mathcal{L}_{distance}
\end{equation}
To illustrate the loss terms, we define the aligned RMSD loss as $\mathcal{L}_{align}(\rmX^{(l)}, \rmX) = \|\text{Align}(\hat{\rmX}, \rmX) - \rmX\|$, given the the ground truth 3D coordinates $\rmX \in \sR^{n,3}$ and the predicted 3D coordinates $\hat{\rmX}=\{\vx_1, \vx_2, \vx_3, \cdots, \vx_n \} \in \sR^{n,3}$. The global, fragment and pair loss are defined by the aligned MSE loss, but with different input data shape:
\begin{itemize}
    \item \textbf{Global Loss}: $\rmX$ with shape $[n,4,3]$. RMSD of the global structure.
    \item \textbf{Fragment Loss}: $\rmX$ with shape $[n,c,4,3]$.  RMSD of $c$ neighbors for each residue.
    \item \textbf{Pair Loss}: $\rmX$ with shape $[n,K,c \cdot 2, 4, 3]$. RMSD of $c$ neighbors for each kNN pair.
    \item \textbf{Neighbor Loss}: $\rmX$ with shape $[n,K,4, 3]$. RMSD of $K$ neighbors for each residue.
\end{itemize}
where $n$ is the number of residues, $c=7$ is the number of fragments, $K=30$ is the number of kNN, $4$ means we consider four backbone atoms $\{N, CA, C, O\}$, and $3$ means the 3D coordinates. The distance loss is defined as the MSE loss between the predicted and ground truth pairwise distances:
\begin{equation}
    \mathcal{L}_{distance} = \|\text{Dist}(\hat{\rmX}) - \text{Dist}(\rmX)\|
\end{equation}
where $\text{Dist}(\rmX) \in \sR^{n,n}$ is the pairwise distance matrix of the 3D coordinates $\rmX$. We apply the loss on each decoder layer, and the final loss is the average, whcih is crucial for good performance.

\section{Experiments}
We conduct systematic experiments to inspire further improvement of FoldToken-V2.
\begin{itemize}%[leftmargin=6.5mm]
   \item{\textbf{AutoEncoder Training (Q1):}} How well the compact and invariant latent space learned by FoldToken2 can be used to reconstruct protein structures?
   \item \textbf{Single-Chain Reconstruction (Q2):} Can FoldToken2 outperform FoldToken1 in single-chain protein reconstruction? 
   \item \textbf{Multi-Chain Reconstruction (Q3):} Can FoldToken2 perform well on multi-chain protein reconstruction?
   % \item \textbf{FoldDiffusion (Q3):} Can FoldToken2 performs well on multi-chain protein reconstruction?
   % \item \textbf{FoldGPT (Q4):} Can FoldToken2 be used to generate protein sequences?
\end{itemize}

\paragraph{Metrics} We evaluate FoldToken2 on the protein structure reconstruction task, where the TMscore and aligned RMSD are reported.

\paragraph{Single-chain Data}  We use the CATH4.3 dataset to train the single-chain model. The same as protein inverse folding models, CATH4.3 is split into training (16631), validation (1516), and testing (1864) sets according to the CAT code. During evaluation, we remove proteins with NaN coordinates, resulting in 493 core samples in the final testing set (T493). Due to the different data representations, backbone architecture and iterative refinement strategies, FoldToken2 trains much slower than FoldToken1. Therefore, we do not train FoldToken2 in the AF2DB dataset like FoldToken1, which can be done in the future if the computing resource is enough. In addition, we also report results over the sub-testing set (T116) of FoldToken1 for head-to-head comparision, which contains 116 proteins.

\paragraph{Multi-chain Data} We train the model using all proteins collected from the PDB dataset as of 1 March 2024. After filtering residues with missing coordinates and proteins less than 30 residues, we obtain 162K proteins for training. We random crop long proteins to ensure that the maximum length is 500. Protein complexes are supported by adding chain encoding features $c_{ij}$ to the edge $e_{ij}$: $c_{ij} = 0, \text{if } i \text{ and } j \text{ are in different chains}$; else $c_{ij} = 1$.
% \begin{equation}
%    c_{ij} = \begin{cases}
%        0, \text{if } i \text{ and } j \text{ are in different chains}\\
%        1, \text{else}
%    \end{cases}
% \end{equation}

\subsection{AutoEncoder Training (Q1)}
\paragraph{Setting} We train FoldToken2 without vector quantization to show how well the encoder-decoder can learn and generate protein structure patterns. The results could be used as the ceil performance of FoldToken series models. We train the model up to 25 epochs with a batch size of 8 and a learning rate of 0.001. The overall model contains 8 encoding layers, 8 decoding layers, and a 128 hidden dimension, total 9.2M parameters. The training process could be done in 1 days using 1 A100 GPU. We also report the structure prediction results of ESMFold \cite{lin2023evolutionary}, AlphaFold2 \cite{jumper2021highly} and AlphaFold3 \cite{abramson2024accurate} for evaluating the structure generation capability of FoldToken2.

\begin{table}[h]
   \centering
   \small
   \begin{tabular}{llllllll}
      \toprule
              & Data & \multicolumn{3}{c}{TMScore}                                                 & \multicolumn{3}{c}{RMSD}                                                    \\
              &      & \multicolumn{1}{c}{Avg} & \multicolumn{1}{c}{Max} & \multicolumn{1}{c}{Min} & \multicolumn{1}{c}{Avg} & \multicolumn{1}{c}{Max} & \multicolumn{1}{c}{Min} \\ \midrule
   FoldToken2 & T493 & 0.98                    & 0.99                    & 0.93                    & 0.46                    & 2.12                    & 0.26                    \\
   ESMFold    & T116 & 0.94                    & 0.99                    & 0.88                    & 1.21                    & 5.33                    & 0.33                    \\
   AlphaFold2 (Colab) &  T116 &  0.92  &  0.99  &  0.35 & 1.44 & 11.00 &  0.38 \\
   AlphaFold3 & T116 & 0.94                    & 0.99                    & 0.74                    & 1.24                    & 4.02                    & 0.33                 \\ 
   FoldToken2 & T116 & \textbf{0.98}                    & \textbf{0.99}                    & \textbf{0.87}                    & \textbf{0.41}                    & \textbf{0.87}                    & \textbf{0.25}                    \\ \bottomrule
   \end{tabular}
   \caption{AutoEncoder reconstruction results of FoldToken2.}
   \label{tab:autoencoder}
\end{table}

\begin{figure}[H]
   \centering
   \includegraphics[width=4.5in]{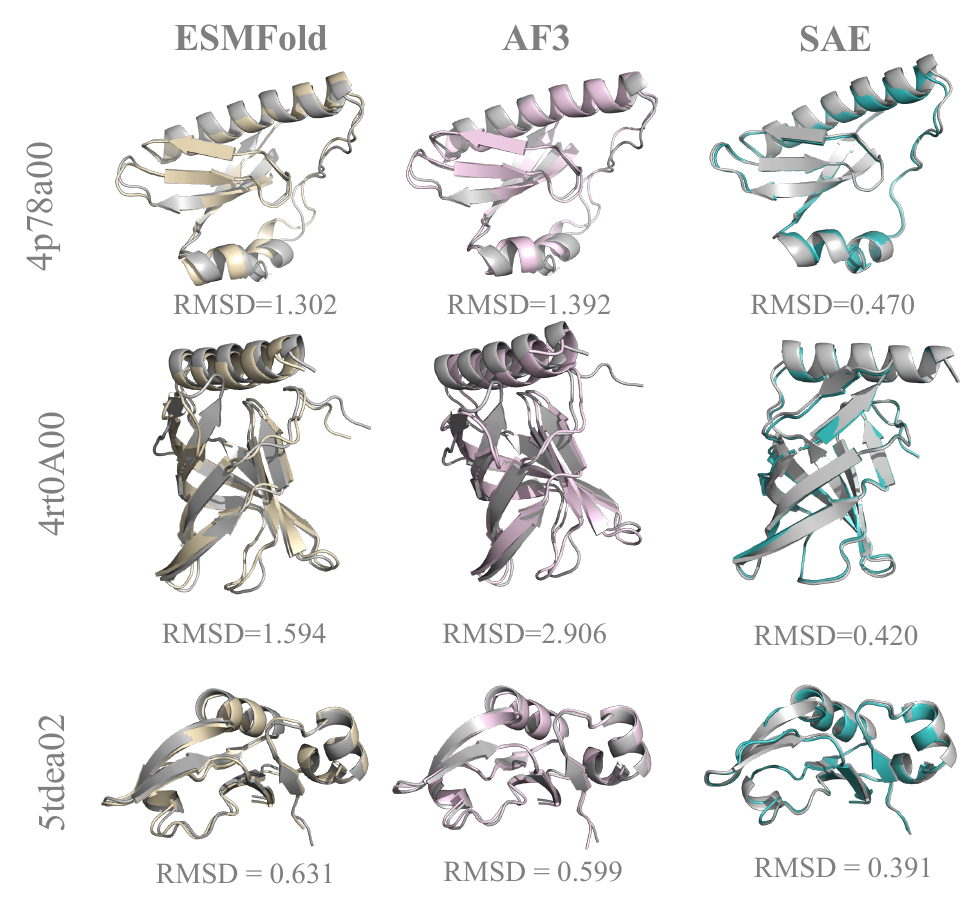}
   \caption{Reconstruction performance without VQ. "SAE" means structure autoencoder without VQ.}
   \vspace{-5mm}
   \label{fig:reconstuct_AE}
\end{figure}

\paragraph{Results} In Table.~\ref{tab:autoencoder}, we show the reconstruction results of FoldToken2.  In both T493 and T116, FoldToken2 achieves good performance in TMScore and RMSD. When compared to ESMFold, AlphaFold2, and AlphaFold3, FoldToken2 can generate more accurate protein structures from invariant latent embeddings, showing the capability of the equivariant decoder. Notabely, the protein folding decoder only contains 4.9M parameters, which is much smaller than the 688.55M parameters of ESMFold's Folding Trunk. FoldToken2 may provide new insights for the protein folding problem.
% encoder+decoder contains 9.2M parameters, and the decoder contains 4.9M parameters

\subsection{Single-Chain Reconstruction (Q2)}
\paragraph{Setting} We compare FoldToken2 with FoldToken1 on both T116 and T493 datasets.  The model is trained up to 25 epochs with a batch size of 8 and leading rate of 0.001. One experiment could be done in 2 days using 1 80G A100 GPU. 

% Please add the following required packages to your document preamble:
% \usepackage{multirow}
\begin{table}[h]
    \resizebox{1.0 \columnwidth}{!}{
    \begin{tabular}{lllllllllllll} 
        \toprule
    \multirow{2}{*}{Model} & \multirow{2}{*}{Data}         & \multicolumn{5}{c}{Config}                                                                                                                          & \multicolumn{3}{c}{TMScore $\uparrow$}           & \multicolumn{3}{c}{RMSD $\downarrow$ }               \\
    &           & \multicolumn{1}{c}{\#Code} & \multicolumn{1}{c}{\#Enc} & \multicolumn{1}{c}{\#Dec} & \multicolumn{1}{c}{\#Hid} & \multicolumn{1}{c}{\#KNN} & \multicolumn{1}{c}{Avg} & Max  & Min  & \multicolumn{1}{c}{Avg} & Max   & Min  \\ \hline
    FT1          & T116         & 65536                           &     12                      &   12                        &     480                      &    --                           & 0.77                    & 0.96 & 0.39 & 3.31                    & 24.53 & 0.52 \\ \cdashline{1-13} \cdashline{1-13}
    \multirow{14}{*}{FT2} & \multirow{14}{*}{T116}  & 65536                           & 6                         & 8                         & 128                       & 30                            & 0.84                    & 0.94 & 0.68 & 1.50                    & 1.98  & 1.16 \\
    & & \chl 65536                           & \chl 8                         & \chl 8                         & \chl  128                       & \chl 30                            & \chl 0.89                    & \chl 0.96 & \chl 0.74 & \chl 1.20                    & \chl 1.68  & \chl 0.88 \\
    & & 65536                           & 10                        & 8                         & 128                       & 30                            & 0.90                    & 0.97 & 0.73 & 1.21                    & 2.57  & 0.55 \\
    &    & 65536                           & 12                        & 8                         & 128                       & 30                            & 0.41                    & 0.73 & 0.25 & 7.51                    & 15.64 & 2.00 \\
    &   & 65536                           & 6                         & 10                        & 128                       & 30                            & 0.82                    & 0.91 & 0.62 & 1.68                    & 2.15  & 1.36 \\
    &   & 65536                           & 6                         & 12                        & 128                       & 30                            & 0.77                    & 0.88 & 0.60 & 2.01                    & 2.91  & 1.59 \\
    &   & 65536                           & 6                         & 15                        & 128                       & 30                            & 0.70                    & 0.84 & 0.52 & 2.48                    & 3.48  & 1.76 \\
    &   & 65536                           & 6                         & 10                        & 256                       & 30                            & 0.30                    & 0.43 & 0.17 & 12.40                   & 22.09 & 5.76 \\
    &   & 65536                           & 6                         & 10                        & 128                       & 50                            & 0.78                    & 0.89 & 0.59 & 1.95                    & 2.62  & 1.44 \\
    &   & 1024                            & 6                         & 10                        & 128                       & 30                            & 0.76                    & 0.87 & 0.57 & 2.34                    & 4.90  & 1.23 \\
    &  & 512                             & 6                         & 10                        & 128                       & 30                            & 0.45                    & 0.74 & 0.28 & 6.94                    & 13.00 & 1.80 \\
    &   & 256                             & 6                         & 10                        & 128                       & 30                            & 0.51                    & 0.86 & 0.29 & 5.69                    & 11.99 & 1.03 \\
    &   & 32                              & 6                         & 10                        & 128                       & 30                            & 0.29                    & 0.45 & 0.18 & 10.02                   & 15.30 & 4.32 \\
    &   & 16                              & 6                         & 10                        & 128                       & 30                            & 0.28                    & 0.41 & 0.16 & 12.88                   & 21.82 & 6.52 \\  \cdashline{1-13} \cdashline{1-13}
    $\text{FT2}^{\dagger}$          & T116         & \crg 65536                           &   \crg  8                      &  \crg 8                        &  \crg   128                      &  \crg  30                           & \crg \textbf{0.97}                   & \crg \textbf{0.98}  & \crg \textbf{0.88} &  \crg \textbf{0.63}                     & \crg \textbf{0.93}  & \crg \textbf{0.42} \\ 
    \bottomrule
    FT1  & T493    & 65536    & 12   &  12   &  480  &  --   & 0.74   & 0.96 & 0.44 & 3.09   & 18.09 & 0.48 \\ \cdashline{1-13} \cdashline{1-13}
   \multirow{14}{*}{FT2} & \multirow{14}{*}{T493}  & 65536                           & 6                         & 8                         & 128                       & 30                            & 0.84                    & 0.95 & 0.61 & 1.63                    & \textbf{4.85}  & 1.10 \\
   &  & \chl 65536                           & \chl 8                         & \chl 8                         & \chl  128                       & \chl 30                            & \chl 0.88                    & \chl 0.97 & \chl 0.68 & \chl 1.38                    & \chl 10.92  & \chl 0.95 \\
   &  & 65536                           & 10                        & 8                         & 128                       & 30                            & 0.86                    & 0.96 & 0.49 & 1.81                    & 17.56  & 0.58 \\
   &  & 65536                           & 12                        & 8                         & 128                       & 30                            & 0.36                    & 0.68 & 0.21 & 9.53                    & 27.33 & 2.77 \\
   &  & 65536                           & 6                         & 10                        & 128                       & 30                            & 0.82                    & 0.94 & 0.56 & 1.68                    & 11.89  & 1.29 \\
   &  & 65536                           & 6                         & 12                        & 128                       & 30                            & 0.77                    & 0.93 & 0.47 & 2.22                    & 7.67  & 1.36 \\
   &  & 65536                           & 6                         & 15                        & 128                       & 30                            & 0.71                    & 0.89 & 0.45 & 2.65                    & 11.46  & 1.95 \\
   &  & 65536                           & 6                         & 10                        & 256                       & 30                            & 0.28                    & 0.41 & 0.12 & 14.30                   & 59.28 & 6.27 \\
   &  & 65536                           & 6                         & 10                        & 128                       & 50                            & 0.78                    & 0.91 & 0.55 & 2.11                    & 7.28  & 1.45 \\
   &  & 1024                            & 6                         & 10                        & 128                       & 30                            & 0.71                    & 0.88 & 0.38 & 3.17                    & 15.21  & 1.29 \\
   &  & 512                             & 6                         & 10                        & 128                       & 30                            & 0.38                    & 0.67 & 0.19 & 9.01                    & 49.81 & 2.58 \\
   &  & 256                             & 6                         & 10                        & 128                       & 30                            & 0.43                    & 0.75 & 0.17 & 11.96                    & 27.52 & 5.20 \\
   &  & 32                              & 6                         & 10                        & 128                       & 30                            & 0.27                    & 0.41 & 0.17 & 11.96                   & 27.52 & 5.20 \\
   &  & 16                              & 6                         & 10                        & 128                       & 30                            & 0.27                    & 0.43 & 0.13 & 14.09                   & 59.69 & 6.18 \\  \cdashline{1-13} \cdashline{1-13}
   $\text{FT2}^{\dagger}$          & T493         & \crg 65536                           &   \crg  8                      &  \crg 8                        &  \crg   128                      &  \crg  30                           & \crg \textbf{0.95}                  & \crg \textbf{0.99}     & \crg \textbf{0.69} &  \crg \textbf{0.86}                     & \crg 12.62  & \crg \textbf{0.41} \\ 
   \bottomrule
    \end{tabular}}
    \caption{Reconstruction performance. We highlight the promising models in color. FT1, FT2, and $\text{FT2}^{\dagger}$ indicates FoldToken1 \citep{gao2023vqpl,gao2024foldtoken}, FoldToken2, and FoldToken2 with the adaptive temperature annealing strategy, respectively. FT2 use the linear temperature decay scheduler, reducing the temperature from 1.0 to $1e^{-5}$ in 40k steps. $\text{FT2}^{\dagger}$ use the adaptive temperature tuner, reducing the temperature from 1.0 to $1e^{-5}$ in 5k steps.}
    \label{tab:reconstruction}
\end{table}

\begin{figure}[H]
   \centering
   \includegraphics[width=4.2in]{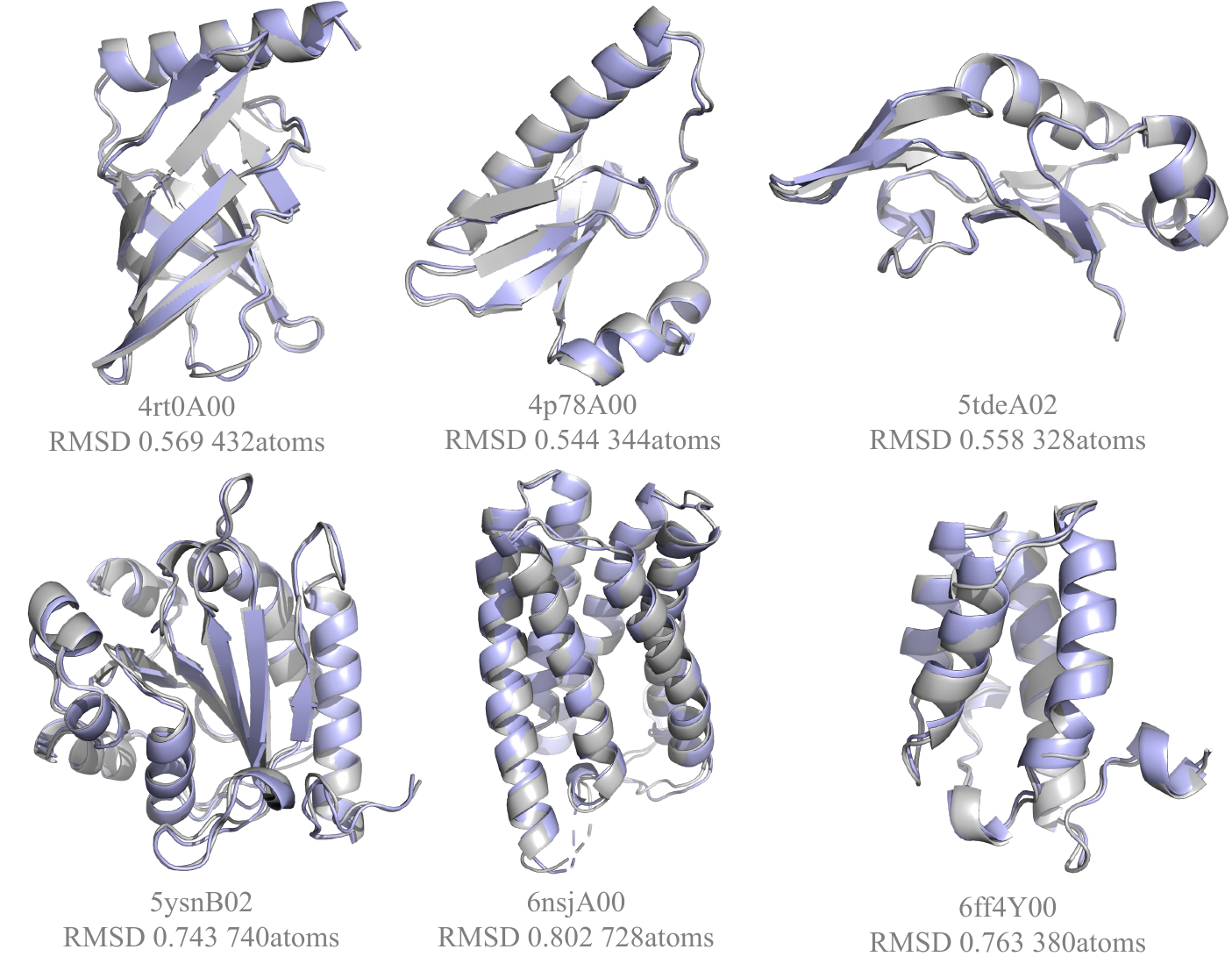}
   \caption{Single-chain reconstruction. Grey and colored residues represent the ground truth and predicted ones.}
   \vspace{-5mm}
   \label{fig:recover_single_chain}
\end{figure}

In Table.~\ref{tab:reconstruction} and Fig.~\ref{fig:recover_single_chain}, we show the reconstruction results of FoldToken2 and conclude that:

\paragraph{FoldToken2 outperforms FoldToken1 by a large margin on structure reconstruction.} In average TMScore, the $\text{FT2}^{\dagger}$ model achieves 0.97 on T116, outperforming FT1 by 20\%. In RMSD, $\text{FT2}^{\dagger}$ achieves 0.63 on T116, outperforming FoldToken1 by 81\%. Similer results are observed on T493.  Regarding the robustness, $\text{FT2}^{\dagger}$ is more reliable, with worst-case TMScores and RMSD of 0.88 and 0.42, respectively, compared to 0.39 and 24.53 for FoldToken1. These results suggest that the new encoder, quantifier, and decoder in FoldToken2 significantly enhance protein structure reconstruction.

\paragraph{Scalling up model does not perform better on limited data.} We find that the 8-layer encoder and decoder with hidden size 128 achieve the best performance. Futher increasing the number of layers or hidden size does not lead to improvement, possiblly limited to the small training data volumn. Also, we should point out that we do not pay much attention to stablize the training of deepper and wider models, and results may improve with careful adjustments. 
\vspace{-3mm}
\paragraph{Improving Decoing KNN do not improve performance.} Increasing the number of KNN neighbors in the decoder from 30 to 50 does not improve the performance. This result suggests that the 30 kNN neighbors are sufficient to capture the structure interactions, providing valuable insights for efficient model design with sparse attention.

% Regarding the number of encoding layers, the model with 10 layers achieves the best average TMScore of 0.90. However, the model with 12 layers performs worse than those with 10 or 8 layers. For the decoder, models with more than 8 decoding layers perform worse than the model with 8 layers. Additionally, models with a 256 hidden dimension perform worse than those with a 128 hidden dimension. 

\paragraph{Larger Codebook Size Leads to Better Reconstruction Performance.} Increasing the codebook size from 16 to 65,536 consistently improves reconstruction performance. This suggests that the 20-codebook size in FoldSeek is likely insufficient for accurately describing protein structures from a reconstruction perspective. However, since FoldSeek \cite{van2024fast} focuses on structure alignment, this loss of information does not significantly affect its widespread usage.

\section{Multi-Chain Reconstruction (Q3)}
\begin{figure}[h]
   \centering
   \includegraphics[width=4in]{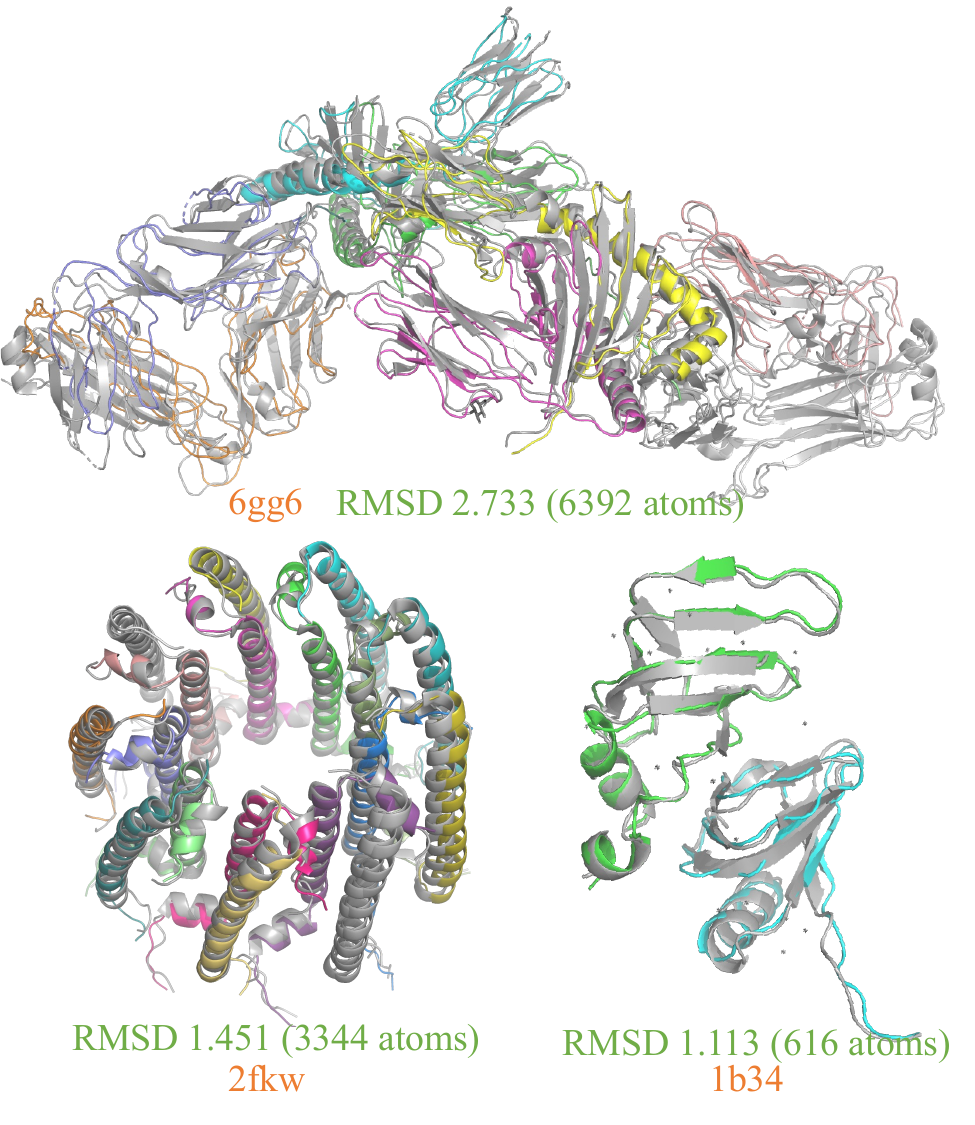}
   \caption{Multi-chain reconstruction. Grey and colored residues represent the ground truth and predicted ones.}
   \vspace{-5mm}
   \label{fig:reconstuct_multi_chain}
\end{figure}

\paragraph{Setting} We train FoldToken2 on the multi-chain protein reconstruction task using the PDB dataset. The model is trained for up to 100 epochs with a batch size of 8, a learning rate of 0.001, and a padding length of 512. There is no available benchmark for evaluating multi-chain protein reconstruction, and we provide some visual examples in Fig.~\ref{fig:reconstuct_multi_chain}. The comprehensive benchmark is coming soon.

\paragraph{FoldToken2 generalize well to large-scale protein systems.} While we crop long proteins to ensure that the maximum length is 500,  FoldToken2 can generate protein complexes with low RMSD error, as shown in Fig.~\ref{fig:reconstuct_multi_chain}. The results suggest that FoldToken2 can be used for multiple real-world applications beyond protein monomers.

\paragraph{We are surprised by the small model size and good training efficiency.} Unlike other models that use AF2 for structure generation, FoldToken2 does not copy AF2. Instead, we use a light-weight model, comprising a 4.31M encoder, 4.27M quantifier, and 4.92M decoder, achieving state-of-the-art results. When generating protein structures, FoldToken2 requires 5000 GPU hours; in contrast, FT2 only requires 40GPU hours training over the whole PDB dataset.

\section{Extension (Future Work)}
\paragraph{TokenFlow} Given a discrete protein language and protein sequence, we train a flow-matching model using rectified flow \cite{liu2022flow} on the CATH4.3 training set. The flow-matching model is designed to predict protein structures from single protein sequences without using MSA information. The model is trained for up to 1000 epochs with a batch size of 128, a learning rate of 0.0005, and a padding length of 512. The overall model architecture consists of a 12-layer transformer, similar to ESM-35M. The training process can be completed in about one day using a single A100 GPU.  When training over 16K data of CATH4.3, the model seems to be not work well as expected. Limited to the energe and computing resource, we plan to scalling the model in the future.

\vspace{-5mm}
\begin{figure}[h]
   \centering
   \includegraphics[width=5in]{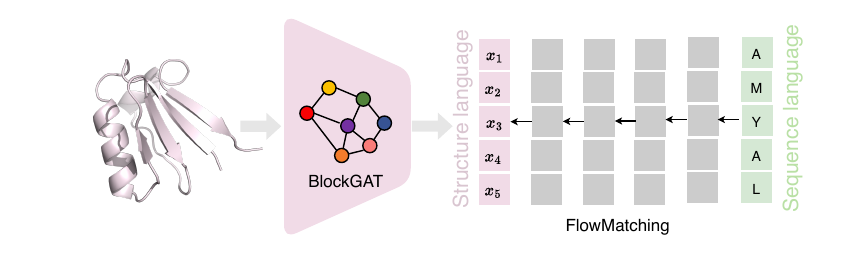}
   \caption{Flow matching for sequence-srtructure translation. }
   \vspace{-5mm}
   \label{fig:TokenFlow}
\end{figure}

\paragraph{FoldGPT} Similar to FoldToken1, we train a model for generating masked structure regions based on unmasked ones. However, the model seems to be overfitting on the training data, and we need to scale up the model in the future.
\begin{figure}[h]
   \centering
   \includegraphics[width=5.5in]{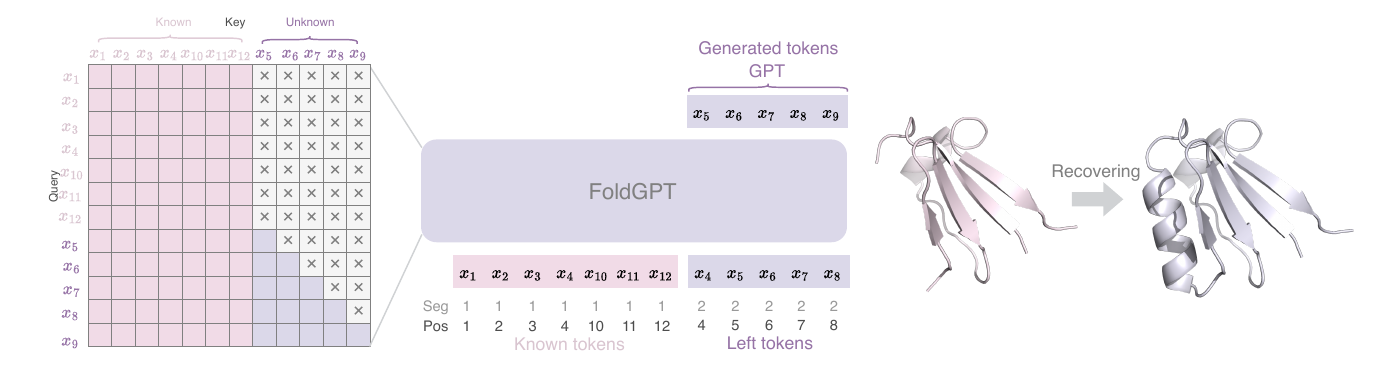}
   \caption{FoldGPT for srtructure generation. }
   \label{fig:FoldGPT}
\end{figure}

\section{Conclusion} 
This paper propose FoldToken2, probably the first multi-chain protein structure tokenization approach. With great efforts in improving the encoder, quantifier, and decoder, FoldToken2 significantly outperforms FoldToken1 by 20\% in TMScore and 81\% in RMSD. Additionally, the training efficiency is remarkable, requiring only 40 GPU hours to train on the entire PDB dataset. The propose methods may inspire further improvement in protein structure representation learning, structure alignment, and structure generation tasks.

\bibliographystyle{plain}
\bibliography{FoldToken2}

% \clearpage
% \appendix
% \section{Appendix}

\end{document}